%% file: acm_main.tex
\pdfoutput=1
\documentclass[sigconf, nonacm]{acmart}

\AtBeginDocument{
  }
\usepackage{subcaption}
\usepackage{graphicx}
\usepackage{booktabs}
\usepackage{siunitx}
\usepackage{xspace}
\usepackage{multirow}
\usepackage[table,xcdraw]{xcolor}
\usepackage{tikz}
\usepackage{tcolorbox}
\usepackage{soul}
\soulregister\cite7
\soulregister\citep7
\soulregister\citet7
\soulregister\ref7
\usepackage{enumitem}
\setlist[itemize]{leftmargin=*, topsep=1.5pt, itemsep=1.5pt}
\setlist[enumerate]{leftmargin=*, topsep=1.5pt, itemsep=1.5pt}

\makeatletter
\AtBeginEnvironment{figure}{\let\@vspace\@vspace@orig\let\@vspacer\@vspacer@orig}
\AtBeginEnvironment{figure*}{\let\@vspace\@vspace@orig\let\@vspacer\@vspacer@orig}
\makeatother

\DeclareRobustCommand{\name}{\textsc{Metadata Reasoner}\xspace}
\soulregister\name0

\setcopyright{acmlicensed}
\copyrightyear{2027}
\acmYear{2027}
\acmDOI{XXXXXXX.XXXXXXX}
\acmConference[SIGMOD '27]{Proceedings of the 2027 International Conference on Management of Data}{June 13--19, 2027}{Huntington Beach, CA, USA}

\begin{document}

\title{An Agentic Approach to Metadata Reasoning}

\author{Jiani Zhang, Sercan \"{O}. Ar{\i}k, Cosmin Arad, Fatma \"{O}zcan, Alon Halevy}
\affiliation{
  \institution{Google}
  \country{}
}
\email{\{jianizhang, soarik, carad, fozcan, halevy\}@google.com}

\renewcommand{\shortauthors}{Zhang et al.}

\begin{abstract}
As autonomous LLM agents evolve to perform complex analytical tasks that require integrating multiple datasets, the problem of discovering relevant data sources becomes a key bottleneck. Beyond the challenge posed by the sheer volume of available data sources, data-source selection is difficult because the semantics of data are nuanced and require considering many aspects of the data. 
To address this challenge, we introduce \name, an agentic approach to metadata reasoning that identifies a small set of data sources that are both sufficient and minimal for a given analytical task. \name leverages a table-search engine to retrieve candidate tables, and then autonomously consults various aspects of the available metadata to determine whether the candidates fit the requirements of the task.
We demonstrate the effectiveness of \name through a series of empirical studies. Evaluated on the real-world KramaBench datasets for data selection, our approach achieves an average F1 score of 88.63\%, outperforming the strongest baseline by 10.73 percentage points. 
Furthermore, evaluations on a newly created synthetic benchmark based on the BIRD dataset reveal that \name is highly robust against redundant and low-quality tables that are in the data lake. In this noisy environment, \name maintains a 93.5\% average F1 score for selecting the right datasets and demonstrates a 100\% success rate in avoiding low-quality data.
Finally, we extend \name to support the interleaving of table selection with the execution of partial queries. Compared with a single pass of table selection, such interleaving improves table-selection F1 score on runtime-dependent tasks by 4.8 percentage points and end-to-end task accuracy by 5.3 points.
\end{abstract}

\maketitle

\input{sections/1_introduction}
\input{sections/2_related_work}
\input{sections/3_metadata_reasoner_architecture}
\input{sections/4_experimental_setup}
\input{sections/5_experimental_evaluation}
\input{sections/5_failure_analysis}
\input{sections/6_agent_integration}
\input{sections/7_conclusion}

\newpage

\bibliographystyle{ACM-Reference-Format}
\bibliography{reference}

\end{document}

%% file: sections/1_introduction.tex
\section{Introduction}

The rapid advancement of autonomous LLM agents has fundamentally transformed data-centric workflows~\citep{fu2025autonomous, zhu2025survey}, enabling unprecedented levels of automation. These agents hold the promise to autonomously execute high-level tasks, such as orchestrating data engineering pipelines~\citep{jin2025elt} or answering business analytics questions~\citep{salemi2025blackboard}. However, a primary bottleneck in developing these agents lies in the effective discovery and selection of data sources that are relevant to a given task~\citep{wei2025revisiting, kosiuk2026fine}. This challenge is especially acute when these agents operate over large-scale repositories in modern data lakes. 

In addition to the sheer number of available data sources, data selection faces two additional challenges. The first challenge is that deciding whether a data source is relevant to a task is nuanced. The schema and textual description of a table are crucial, but many other factors need to be considered, including the quality and recency of the data, whether it is governed, how it was created (including its code and data lineage), the purpose it was created for, and hidden assumptions that went into its creation. While present-day data catalogs can accommodate many of these aspects, they are built for humans. They rely on rich interfaces and implicit context that a human analyst can effortlessly interpret, yet this context remains opaque to an LLM agent. 

The second challenge is that complex real-world agentic tasks typically require a combination of data sources rather than a single one. In doing so, it is necessary to guarantee that the collection of data sources together covers all the entities and attributes needed for the task, and that the data sources can be combined (e.g., via joins or unions) to yield the correct result.

One might assume that the expanding context windows of modern LLMs would solve this by allowing agents to ingest vast amounts of metadata and reason over it directly~\citep{chung2025long, comanici2025gemini, yu2025memagent}. However, performance often degrades with context length due to attention dilution across extended contexts, which causes the model to overlook crucial schema constraints~\citep{liu2024lost,shi2023distracted,li2024retrieval}. Furthermore, the scale of enterprise data lakes makes reasoning over the entire metadata repository computationally prohibitive.

To manage the sheer scale of available data sources and their associated computational costs, the de facto approach to data discovery today utilizes ranking-based retrieval systems~\citep{brickley2019googledatasetsearch, balaka2025pneuma}. These systems may integrate multiple retrieval signals: a lexical full-text search (such as BM25~\citep{robertson2009probabilistic}) to handle exact string matches, a semantic vector search that encodes the descriptions of data sources into an embedding space to capture their semantics~\citep{malkov2018hnsw}, and a re-ranking mechanism to refine the final ranking. 
However, because these methods are inherently designed to score and retrieve datasets individually, they remain inadequate for retrieving combinations or integrations of multiple data sources needed for a task.
Consider queries requiring attributes $A_1$ and $A_2$ from distinct tables: a ranking-based search may return sources highly relevant to $A_1$ while omitting any sources relevant to $A_2$. Even if the search does find tables covering both attributes, these tables may not be connected, either because they contain different sets of entities or because an intermediate table is needed as a relational bridge. Furthermore, when a join path exists, determining whether it is semantically valid (for instance, distinguishing between a \emph{customer} table joined to \emph{orders} versus \emph{returns}) requires reasoning about the intent of the relationship, not just the presence of a shared key. Ranking models optimize for semantic relevance but are not designed to capture other aspects of metadata reasoning. 

This paper introduces \name, an agentic approach to metadata reasoning designed to identify a small set of data sources that are both sufficient and minimal for a given analytical task. \name takes as input an analytical task and a comprehensive metadata catalog that stores all known properties of the available data sources. This catalog captures not just the schemas and text descriptions of tables, but also their creation history, data quality metrics, usage patterns, and broader business definitions such as glossaries or ontologies. 
 
Rather than bringing all available metadata into the LLM's context window, \name executes a series of steps that involve retrieving candidate tables and reasoning about both their attached metadata and additional on-the-fly metadata when needed. First, to retrieve candidate tables effectively in large data lakes, where tables with similar schemas or topics often produce nearly indistinguishable embeddings because subtle differences occupy only a small fraction of each table's metadata, we introduce a two-stage discrimination-oriented description generation pipeline that groups related tables and contrasts them to highlight their distinguishing attributes, such as temporal ranges. Second, once candidate tables are retrieved, \name reasons over their attached metadata, which provides a concise summary of each table. Third, when attached metadata leaves ambiguities, \name uses specialized tools to fetch on-the-fly metadata, such as checking whether two tables can be meaningfully joined, verifying whether a particular value is present in a given table, or consulting a business glossary to better understand the intent of the task.
\name autonomously decides which metadata to consult at any given point, depending on how it is progressing towards its goal. 

We evaluate \name on two benchmarks: KramaBench~\citep{lai2025kramabench}, a real-world benchmark representing a messy data lake across six domains, and a synthetically scaled-up version of the BIRD dataset~\citep{li2023bird}. 
We select KramaBench because it challenges systems with high-cardinality search spaces (approximately 1,500 candidate tables in the Astronomy domain) and requires multi-source reasoning.
To systematically stress-test the system's resilience against the complexities of modern enterprise data environments, we expand the BIRD dataset by injecting horizontal partitions, redundant duplicates, and structural noise.
We implement \name in two ways: one as a dedicated agent, which owns the reasoning loop, and the other as a harness plugin, with which the host agent of an agent harness performs the reasoning itself, calling each metadata fetch and reasoning operation as a separate tool or skill.

Our empirical results show that \name substantially improves table selection accuracy while keeping reasoning overhead low. On KramaBench, \name reaches up to 88.63\% average F1 score, outperforming the strongest baseline, Agentic SQL, by up to 10.73 percentage points and more than doubling vector search accuracy in a domain with over a thousand tables. In the noisy synthetic BIRD data lakes, \name maintains a 93.5\% average F1 score, compared to 30.0\% for vector search, and completely avoids selecting corrupted or low-quality tables. Furthermore, discrimination-oriented descriptions boost retrieval Recall@5 to 70.71\%, while combining attached statistical table descriptions with specialized tools for on-the-fly metadata reduces tool-probing steps by up to 43\%, enabling \name to converge in 8 to 10 steps on average.

Finally, in many situations, it is not possible to find all the relevant tables ahead of time because some metadata decisions can only be made once part of the query has been executed. We extend \name to interleave planning and execution. We consider multiple ways of integrating \name with a host agent, and find that interleaving planning and execution by letting \name request intermediate runtime bindings from the host agent performs best, improving the table-selection F1 score on runtime-dependent tasks from 79.9\% to 84.7\% and overall end-to-end task accuracy from 79.0\% to 84.3\% on runtime-dependent tasks over single-pass selection.

In summary, this paper makes the following contributions:
\begin{itemize}
    \item \textbf{Agentic Metadata Reasoning}: We introduce \name, an agentic approach that combines semantic search, attached statistical metadata summaries, and specialized tools and skills that fetch on-the-fly metadata when necessary while verifying coverage, quality, and joinability.
    \item \textbf{Discrimination-Oriented Table Description Generation}: To separate similar tables during vector search, we design a two-stage pipeline that groups related tables and contrasts them to highlight the differences, which improves retrieval recall over raw schemas and single-table summaries.
    \item \textbf{Dedicated Agent and Harness Plugin Implementations}: We implement \name both as a dedicated agent and as a modular harness plugin equipped with tools, agent skills, and behavioral rules, and evaluate both implementations, showing that while both surpass all baselines, the Harness Plugin outperforms the Dedicated Agent in average F1 by 1.54 percentage points with Gemini 3.8 Flash and 4.92 percentage points with Claude Sonnet 5.
    \item \textbf{Interleaving Planning and Execution}: We evaluate three integration strategies with a code-executing host agent, demonstrating that letting \name request runtime bindings from the host agent overcomes the limits of single-pass table selection and produces more precise handoffs on runtime-dependent tasks.
\end{itemize}

%% file: sections/2_related_work.tex
\section{Related Work}
\label{sec:related_work}

\subsection{Dataset search and data integration}
Dataset search aims to fulfill information needs across vast collections like data lakes~\citep{chapman2020dataset,hulsebos2024took, leventidis2024large,tablediscovery, fernandez2018aurum}.
While early keyword-based methods struggle with semantic ambiguity~\citep{dataexplorationsurvey, brickley2019googledatasetsearch},
subsequent approaches identify joinable and unionable tables~\citep{koutras2021valentine, nargesian2018table} utilizing approximate containment search~\citep{bogatu2020D3L, zhu2019josie}, knowledge graphs~\citep{koutras2025omnimatch}, or representation learning~\citep{grace2023starmie}.
To match schema elements under context constraints, recent work proposes structure-guided LLM frameworks for context-aware, budgeted schema matching~\citep{chen2026construm}.
LLMs have also been applied to data discovery and integration~\citep{freire2025large}, including tabular representation and retrieval~\citep{balaka2025pneuma}, data lake discovery~\citep{an2025ledd}, task-oriented search~\citep{wei2025revisiting}, automated metadata generation~\citep{zhang2025autoddg}, and interactive query reformulation~\citep{lin2025rethinking}.
To handle complex queries, recent work has shifted towards fine-grained table retrieval by decomposing typed queries for better connectivity-awareness~\citep{kosiuk2026fine} or employing hierarchical LLM reasoning to retrieve precise sub-tables~\citep{sun2026fgtr}. For evaluation, benchmarks have been introduced to formalize the need for instruction-following in table retrieval~\citep{jin2026followtable}, alongside generative approaches that synthesize complex tabular corpora to stress-test discovery methods~\citep{dai2025lakegen}.

Concurrently, classical data integration systems~\citep{DBLP:books/daglib/0029346} rely on formal descriptions of data sources, where query translation is performed by specialized algorithms (e.g., answering queries using views~\citep{DBLP:journals/vldb/Halevy01}). However, formal descriptions cannot capture the semantic nuances of real-world data lakes. Unlike these systems, \name accepts nuanced natural language descriptions of data sources and reasons about their suitability. Classical database systems also study how source selection and query planning interact with runtime execution. In data integration over sources with limited access patterns~\citep{rajaraman1995answering}, a relation with a \emph{binding pattern} requires specific input attributes to be bound before it can return tuples, so a query plan executes an initial subplan to obtain those bindings before accessing the dependent source. Similarly, adaptive query processing~\citep{DBLP:journals/ftdb/DeshpandeIR07} interleaves planning with execution and re-invokes the optimizer as intermediate results become available, rather than committing to a single static plan upfront.

\subsection{Retrieval within agentic systems}
Agentic search~\citep{singh2025agenticrag} is an iterative retrieval paradigm that uses LLM reasoning and tool execution to dynamically plan, run, and refine multi-step queries for data retrieval~\citep{zhang2024agenticir, su2024bright, long2025diver, shao2025reasonir, das2025rader}.
This paradigm enables autonomous data agents to manage and explore data lakes~\citep{fu2025autonomous, zhu2025survey}.
For data lakes, recent systems use ``relational reification'' to interactively align agent search with user needs via schema refinement~\citep{PneumaSeeker2026}.
However, recent benchmarks demonstrate that exploratory question answering over data lakes remains challenging, requiring intensive search and reasoning across multiple sources~\citep{wang2026lakeqa}.
\name distinguishes itself from these approaches in several key aspects. While existing frameworks often rely on interactive user alignment~\citep{PneumaSeeker2026}, \name focuses on autonomous execution.
Moreover, instead of ranking individual tables or processing unstructured text, \name reasons over rich metadata catalogs to resolve dependencies (such as joinability and content search) and selects multi-source combinations that are both sufficient and minimal for the task.
General agentic frameworks that interleave reasoning and acting~\citep{yao2022react} struggle in large enterprise data lakes. Naive tool-calling loops or raw SQL queries against system metadata (e.g., \texttt{INFORMATION\_SCHEMA}) cause context window saturation and high token costs.

\subsection{Improving LLM agents}
As LLMs have evolved to interact with external tools~\citep{schick2023toolformer}, a major bottleneck has been context window saturation caused by providing exhaustive API documentation.
Recent studies address this by abstracting verbose documentation into concise instructions~\citep{yuan2025easytool} or employing structured reasoning-and-observation loops~\citep{zhuang2024toolnet, liu2024controlllm}. Inspired by these advances, \name translates high-level tasks into coordinated retrieval steps across diverse metadata catalogs.
Furthermore, despite larger LLM context windows, models are easily distracted by irrelevant context~\citep{shi2023distracted} and perform worse when critical information is buried in long prompts~\citep{liu2024lost}, especially in database-grounded tasks~\citep{chung2025long}. This limitation motivates \name's core objective: identifying a \emph{minimal} yet sufficient set of data sources to avoid distracting the downstream execution agent.
While tool-use frameworks like the Model Context Protocol (MCP) allow agents to interact with external servers, recent agent harnesses also support modular \emph{agent skills}~\citep{agentskill} that guide the host agent's workflow. We demonstrate that \name supports both paradigms, operating effectively either as a dedicated agent invoked via MCP or as a harness plugin composed of reusable agent skills and tools.

%% file: sections/3_metadata_reasoner_architecture.tex
\begin{figure}[tb]
    \centering
    \includegraphics[width=\linewidth]{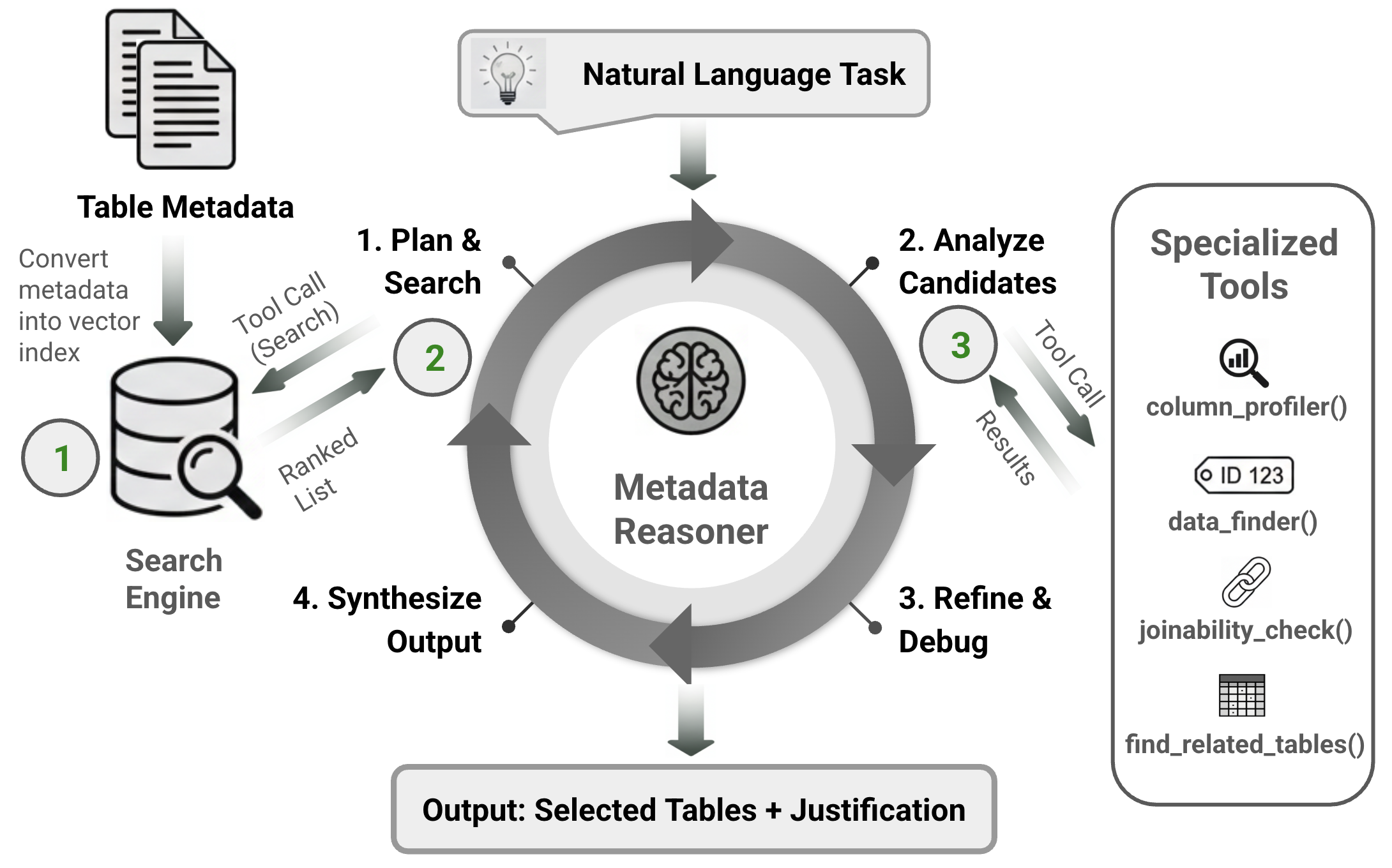}
    \caption{The architecture of \name. Given an analytical task and a metadata catalog, the agent autonomously executes a series of planning, search, and reasoning steps. It leverages both attached metadata from retrieved candidate tables and on-the-fly metadata accessed via specialized tools to output a sufficient and minimal set of tables, along with a natural-language justification. \name runs either as a dedicated agent or as a plugin of an agent harness.}
    \label{fig:metadata_reasoner}
\end{figure}

\section{The Architecture of \name}
\label{sec:methodology}

We first formally define the goal of \name and then describe its architecture and underlying design rationale. 

\subsection{Problem formulation}
\name accepts two \textbf{inputs}:

\smallskip
\noindent
{\bf A data task $Q$:} A typical example of a task is a question that can be answered by a SQL query, or a machine learning prediction or root-cause analysis task that requires Python code.

\smallskip
\noindent
{\bf A metadata catalog $\mathcal{T}$:} The catalog encapsulates all known properties of the available data sources within the data lake. While we use the term \emph{tables} for simplicity throughout this paper, these resources can also encompass collections of structured and semi-structured datasets, such as those found in web tables and open data portals. The catalog captures both structural and semantic information, including schemas and textual descriptions of tables and their attributes. It also includes information surrounding the tables, such as their lineage, quality, and usage patterns. The catalog also contains semantic information that is not attached to a specific table. For example, it may contain a glossary that explains business terms in the domain of the enterprise, an ontology that describes classes of entities and relationships between them, or a semantic model of the data.

\name produces two \textbf{outputs}:
The first is \textbf{a subset of tables} $T^* \subseteq \mathcal{T}$. $T^*$ satisfies the following conditions: (1) \textbf{sufficiency} (or coverage): $T^*$ should contain tables that are required to solve the input task (e.g., tables for answering a query). Specifically, this means that the tables in $T^*$ should cover all the entities and attributes relevant to $Q$, while ensuring that the tables can be joined correctly when needed, and (2) \textbf{minimality}: $T^*$ should be minimal, in the sense that no strict subset $T' \subsetneq T^*$ is sufficient for $Q$. This ensures that every individual table included in $T^*$ is \textbf{necessary} to solve the task.

\noindent \name also outputs \textbf{a natural-language justification} $J$, which provides evidence supporting the sufficiency and minimality of $T^*$ so that downstream agents understand why and how each selected table should be used.

\begin{figure}[tb]
    \centering
    \includegraphics[width=\linewidth]{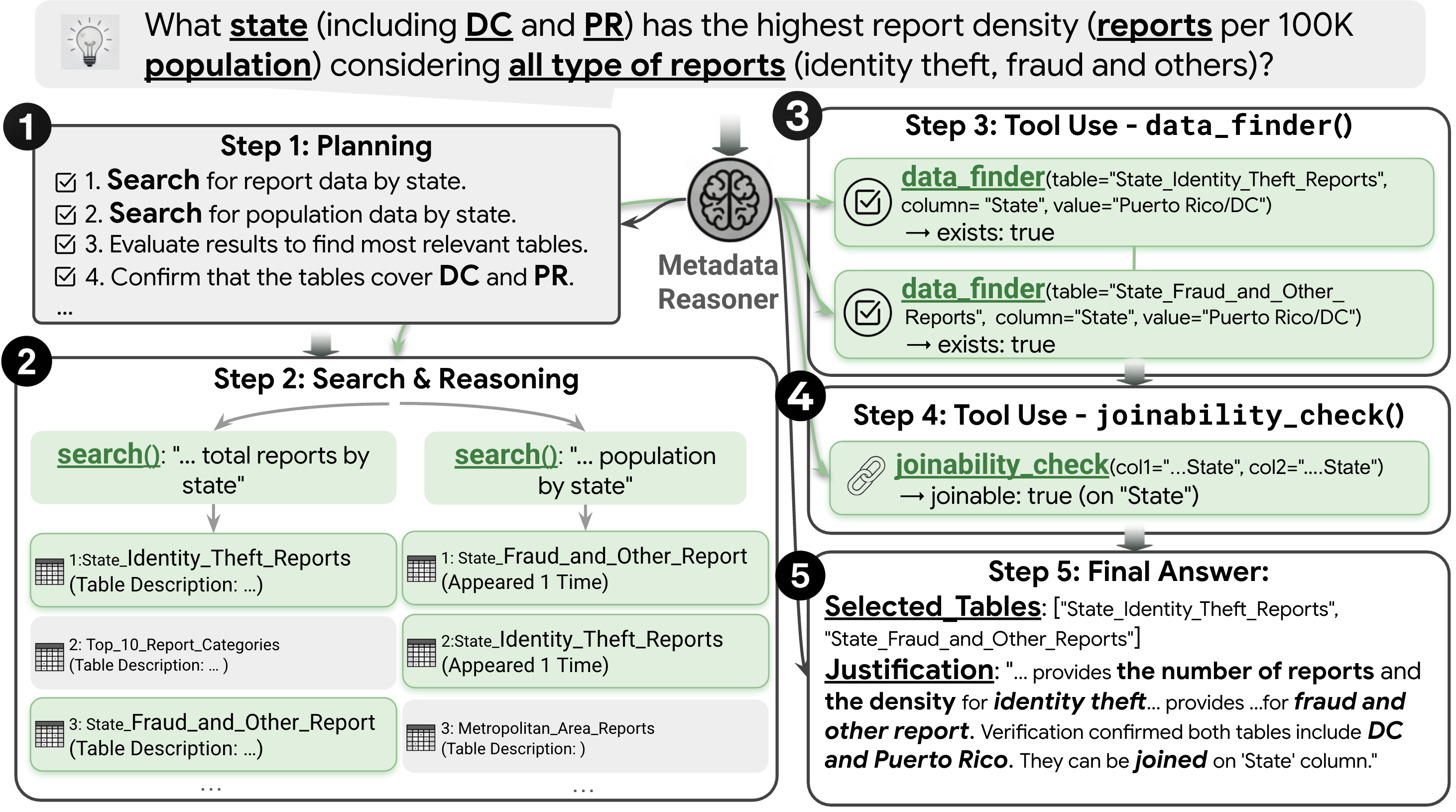}
    \caption{Illustrative example of the \name workflow. Given an analytical task, \name first decomposes the query into multi-faceted search plans (e.g., reports vs.\ population) to retrieve candidate datasets. It then uses specialized tools to confirm entity existence (e.g., ``Puerto Rico/DC'') and to validate relational paths. The final output provides a justified selection of tables, ensuring they are both sufficient for the task and can be joined.}
    \label{fig:mrexample}
\end{figure}

\begin{example}
Figure~\ref{fig:mrexample} illustrates the \name workflow. The agent begins by analyzing the given query to understand which entities (e.g., \emph{DC} and \emph{PR}) and attributes (e.g., \emph{total reports} and \emph{population}) are necessary and what other constraints (e.g., \emph{all types of reports}) apply. Based on this analysis, the agent creates an initial plan for its reasoning. In the example, the first two steps of the plan invoke the search engine to retrieve candidate datasets for \emph{state reports} and \emph{population}. It then uses \texttt{data\_finder()} to confirm the presence of \emph{DC} and \emph{PR} and \texttt{joinability\_check()} to validate that disparate report categories can be unified via shared \texttt{State} columns. If the candidates are insufficient or lack a valid join path, \name recursively reformulates its search strategy to expand the set of candidate tables. 
Upon satisfying all constraints, the agent terminates and outputs a set of selected tables and a natural-language justification for their relevance. In this instance, the reasoner clarifies how the \texttt{State\_\allowbreak Identity\_\allowbreak Theft\_\allowbreak Reports} and \texttt{State\_\allowbreak Fraud\_\allowbreak and\_\allowbreak Other\_\allowbreak Reports} tables satisfy the ``all types'' requirement, ensuring both sufficiency (coverage of all attributes) and minimality (no smaller subset satisfies the constraints).

\end{example} 

\subsection{An agentic approach to metadata reasoning}
As illustrated in Figure~\ref{fig:metadata_reasoner} and the preceding example, \name uses metadata in three ways. First, to retrieve candidate tables, the search engine encodes table descriptions as vector embeddings. Second, once candidate tables are retrieved, the agent reasons over their \emph{attached metadata}, which provides a concise summary of each table. Third, when attached metadata leaves ambiguities, the agent uses specialized tools to fetch \emph{on-the-fly metadata} that would be impractical to provide ahead of time.

\name is built on a core principle: prepare each type of metadata in the form that makes it most useful, and let the agent decide when and where to use it. Supplying all catalog metadata at once is computationally prohibitive and saturates the context window, and hard-wiring which metadata each step consumes cannot anticipate the ambiguities that a particular task raises. Instead, \name exposes each type of metadata through a specialized tool and leaves it to an orchestration agent to fetch and reason over that metadata as the need arises. Given a query $Q$, the agent decomposes it into search constraints, such as named entities, measures, temporal scopes, and granularities, to form a search plan. It then expands its search until the candidate tables cover all constraints, invokes tools for on-the-fly metadata when attached metadata leaves ambiguities, and prunes unneeded tables to return $T^*$. We detail each component below.

\subsection{The semantic search tool}
\label{sec:semantic_search_engine}
The agent retrieves candidate tables from the catalog through a semantic search tool. The tool accesses a set of vectors, each representing the embedding of metadata about a table, and retrieves candidates via approximate nearest-neighbor search over the query embedding. 
A central challenge is deciding which metadata to encode into the table embeddings. In large-scale data lakes, tables that share similar schemas or domain topics often suffer from semantic homogenization, where their embeddings become nearly indistinguishable and retrieval recall degrades because subtle distinguishing details occupy a small fraction of their descriptions.

\subsubsection{Discrimination-oriented metadata construction}
\label{sec:method_discriminative_description}

To support targeted search, we design the embeddings to emphasize the differences between related tables. Specifically, we prioritize unique attributes such as specific temporal ranges, spatial coverage, or partitioning keys. In addition, we ensure that technical identifiers and cryptic column names are mapped to standardized domain terminology and synonymous business logic. This helps align natural-language queries with the indexed vector embeddings of the metadata. We construct these descriptions via a two-stage LLM pipeline.

In the first stage, we group tables based on semantic relatedness. Grouping can be derived from table descriptions or, as in KramaBench, from an existing directory hierarchy.
We task the LLM with identifying shared and unique variables across each group and generating a prompt template that highlights each table's distinguishing features relative to its peers.
In the second stage, we apply the prompt template to each table in the group to produce its differentiated description.
For example, consider the discrimination-oriented descriptions generated for the two tables below. While these tables share an identical schema, they differ in temporal ranges, so our method creates a prompt template that preserves shared domain context (e.g., ``NOAA geomagnetic Kp index forecast'') while foregrounding each table's specific dates.

\begin{tcolorbox}[left=1pt, right=0pt, top=1pt, bottom=1pt, fontupper=\footnotesize]
\begin{verbatim}
Table: 0309geomag_forecast
"NOAA geomagnetic Kp index forecast specifically for March 10, 
March 11, and March 12. The data is recorded in 3-hour intervals and 
consists of floating-point values representing the predicted 
geomagnetic activity levels for each day."

Table: 0310geomag_forecast
"NOAA geomagnetic Kp index forecast specifically for March 11,
March 12, and March 13. The data provides floating-point forecast 
values at 3-hour intervals in Universal Time (UT)."
\end{verbatim}
\end{tcolorbox}

\subsubsection{Handling duplicate search results}

When the agent performs multiple calls to the semantic search tool, some tables may be returned across multiple search queries. Repeating the full metadata for these tables would pollute the context window and cause the agent to overlook distinct tables surfaced in later iterations.

To address this challenge, we implement a state-aware deduplication mechanism within the search tool.
The mechanism maintains a session-level dictionary $\mathcal{S}$ that tracks the identifiers and retrieval counts of all previously surfaced tables. For each retrieved table $\mathit{ID}$, if $\mathit{ID} \notin \mathcal{S}$, the table's full metadata is added to the context window. If $\mathit{ID} \in \mathcal{S}$, the metadata is replaced by a concise recurrence indicator (e.g., ``Table ID: xxx (Appeared $N$ times)'') and its counter in $\mathcal{S}$ is incremented.
This strategy prevents repetitive context bloat and directs the agent toward previously unseen data sources. When a search cycle yields only duplicates, the tool returns a termination signal that prompts the agent to revise its search query.

\subsubsection{Retrieving related tables}
\label{sec:related_tables}
Data lakes often split one dataset into many tables that share a schema, for example one table per year or per state. Because discrimination-oriented descriptions distinguish these tables, a search typically returns only the table that best matches the query, even when the task needs several of them. We therefore complement the search tool with a \texttt{find\_\allowbreak related\_\allowbreak tables()} tool. Given a retrieved table, this tool retrieves sibling tables that share an identical or compatible schema (often accompanied by a common naming stem), as well as other tables co-located in the same dataset, together with their attached metadata.

\subsection{Reasoning about retrieved tables}
Once the search engine retrieves candidate tables, the agent reasons about their suitability for the task $Q$. Here we choose which metadata about a table $T$ to attach directly when $T$ is retrieved (\emph{attached metadata}) versus which metadata to expose on demand through tools (\emph{on-the-fly metadata}). 

Several types of metadata cannot be statically attached to a single table. First, resources such as business glossaries or domain ontologies apply across the entire data lake rather than to a single table.
Second, relational properties such as join paths across table pairs are impractical to precompute exhaustively across a large repository (even though some catalogs may store the most common join paths), and detailed column value distributions are too large to inline without saturating the context window.

We therefore equip \name with specialized tools that compute on-the-fly metadata on demand, and instruct the agent to invoke these tools whenever the attached metadata is insufficient to resolve a constraint. The balance between attached and on-the-fly metadata depends on metadata availability, summary size, latency budgets, and the agent's tool orchestration capacity.

\subsubsection{Attached metadata}
\label{sec:tier2_metadata}
The attached metadata enables the agent to evaluate whether a candidate table satisfies the requirements of $Q$, including its schema, temporal and spatial scope, granularity, and foreign-key references:

\noindent \textbf{Schema Information}: The physical schema of each candidate table, including the table name, column names, data types, constraints, and any existing human-authored documentation.

\noindent \textbf{Semantic Enrichment via Statistical Synthesis}: A data profiler computes descriptive statistics for each table offline, including value ranges, top categorical values, and null ratios. An LLM synthesizes these statistics into concise natural-language column descriptions ahead of time on the catalog $\mathcal{T}$, providing rich semantic grounding without inflating prompt length.

An attached metadata example is shown below.
Whereas the discrimination-oriented description notes generally that the table contains floating-point forecast values, the content-driven attached profile supplies concrete statistical bounds:

\begin{tcolorbox}[left=1pt, right=0pt, top=1pt, bottom=1pt, fontupper=\footnotesize]
\begin{verbatim}
Table: 0309geomag_forecast
"""Table Description: NOAA Kp index forecast 10 Mar - 12 Mar
This table provides a geomagnetic forecast from March 10th to March 
12th. Each row corresponds to a specific time interval. The columns 
represent the forecasted geomagnetic activity for each day.

Schema:
- Time_UT: STRING (NULLABLE) - "...time ranges, e.g., '00-03UT'."
- Mar_10: FLOAT (NULLABLE) - "Forecasted geomagnetic activity value 
for March 10th...ranging from 3.0 to 5.0, and an average of 3.75..."
- Mar_11: FLOAT (NULLABLE) - "...for March 11th...from 3.33 to 3.67."
..."""
\end{verbatim}
\end{tcolorbox}

\subsubsection{On-the-fly metadata}
\label{sec:verification}
Alongside \texttt{find\_related\_tables()}, we provide the following tools to fetch on-the-fly metadata:
\begin{itemize}
    \item \texttt{column\_\allowbreak profiler()}: Invoked when summarized table descriptions lack the detail needed for final data selection. It computes exact statistics for one or more columns, such as row, distinct, and null counts and value ranges. It also returns all values of columns with few distinct values and all rows of small tables.
    \item \texttt{data\_finder()}: Value-level grounding is a query-time requirement, as the specific entities of interest (e.g., unique user IDs or country codes) are only known once the query is received and candidate tables are established. This tool returns lightweight, deterministic boolean evidence of a value's presence by running a targeted SQL query, avoiding token-heavy raw data dumps. One call can check several candidate tables and columns to identify which table contains an entity, and also reports near matches when a value is stored in a different format (such as a longer site name or prefix).
    \item \texttt{joinability\_check()}: On-demand joinability analysis is essential because providing all pairwise joinability information to the LLM in advance is computationally wasteful. Instead, when the agent deems two tables necessary, the tool measures the overlap (intersection and directional containment ratios) and uniqueness of their join keys in the data.
\end{itemize}

The agent decides whether and when to invoke each tool based on its current reasoning needs.

\subsubsection{Pruning}
\label{sec:pruning}
Before returning $T^*$, \name enforces minimality through an explicit necessity check: for each candidate table, it verifies what required query attributes or entities are uniquely provided by that table, and prunes any table whose contribution is subsumed by the remaining candidates. The output justification $J$ documents the unique role of each retained table.

\subsection{Framework independence}
The method described in this section does not depend on a specific agent framework. The semantic search tool, the attached and on-the-fly metadata, and the selection criteria can be packaged either as a dedicated agent that other agents invoke as a single tool or as a harness plugin with which the host agent of an agent harness performs the reasoning itself. Section~\ref{sec:implementations} describes the two implementations we evaluate and their configurations.

%% file: sections/4_experimental_setup.tex
\section{Experimental Setup}
\label{sec:exp_setup}

This section details our evaluation datasets, synthetic data generation pipeline, baselines, implementations, and evaluation metrics.

\subsection{Datasets}
We evaluate \name using two primary benchmarks to capture distinct enterprise data challenges.

\begin{table}[tb]
\centering
\caption{KramaBench dataset statistics. ``Avg.\ \#col.'' stands for the average number of columns per table. ``Avg.\ \#tables'' means the average number of ground-truth tables per task.}
\label{tab:kramabench_stats}
\setlength{\tabcolsep}{4.5pt}
    \begin{tabular}{l | rr | rr}
    \toprule
    \textbf{Domain} & \textbf{\#Tables} & \textbf{Avg.\ \#col.} & \textbf{\#Tasks} & \textbf{Avg.\ \#tables}  \\ 
    \midrule
    Astronomy   & 1498  & 28.6 & 12 & 17.2  \\ 
    Legal       & 134   &  2.5 & 30 & 16.8  \\ 
    Environment & 36    & 10.3 & 20 & 7.9   \\ 
    Biomedical  & 29    & 35.4 & 9  & 2.0   \\ 
    Wildfire    & 21    & 12.1 & 21 & 1.7  \\ 
    Archaeology & 5     & 16.2 & 12 & 1.5   \\ 
    \bottomrule
    \end{tabular}
\end{table}

\subsubsection{KramaBench}
To evaluate performance in real-world settings, we use KramaBench~\citep{lai2025kramabench}, a benchmark representing data lake environments across six domains: astronomy, legal, environment, biomedical, wildfire, and archaeology. While KramaBench originally evaluated end-to-end analytical workflows, we adapt its analytical tasks and reference data sources to focus on the data discovery phase. This benchmark challenges autonomous agents by requiring cross-table analytical reasoning and navigation of high-cardinality search spaces (e.g., approximately 1,500 tables in the Astronomy domain).

The original data sources are organized as directories of heterogeneous files (e.g., raw CSVs, spreadsheets, or specialized scientific formats). To simulate an enterprise data warehouse, we transform these files into structured tables by recursively traversing the directories and applying file-specific parsers (e.g., fixed-width \texttt{.tle} orbital data or \texttt{.omni2} headers) to extract tabular data and descriptive text. We standardize the extracted tables into a schema-aware CSV format and validate column names for database compatibility. We consolidate unstructured descriptions, whether embedded in file headers or located in separate \texttt{.txt} metadata files, into the table description field.
Table~\ref{tab:kramabench_stats} summarizes the dataset characteristics, illustrating the scale and complexity of each domain.

\subsubsection{Synthetically scaled BIRD}
To stress-test system resilience against enterprise data complexity, we synthetically created a data lake containing structurally similar tables with varying data quality. We based this generation on the BIRD dataset~\citep{li2023bird}, which provides natural language questions paired with clean database schemas and ground-truth SQL queries. When evaluated on clean base BIRD schemas, \name achieves near-perfect table selection F1 scores (averaging 95.8\%), demonstrating that table selection over small, curated databases fails to measure the large-scale data discovery bottleneck.
We created two versions of five BIRD databases: a clean version containing only the original base tables, and a messy version containing redundant views, fragmented tables (horizontal partitions), and tables with degraded data quality. Table~\ref{tab:bird_synth} summarizes the statistical characteristics and scale of both versions. We evaluated data selection performance using 758 analytical questions associated with these five databases.

\begin{table}[bt]
    \centering
    \caption{Synthetic BIRD data lake statistics. ``Dup.'' stands for duplicates, and ``Low Q.'' stands for low-quality tables.}
    \setlength{\tabcolsep}{3.5pt}
    \begin{tabular}{c
    | r r r r r | r}
    \toprule
         & Base & Splits & Dup. & Low Q. & \#Tables & \#Tasks\\
        \midrule
        Superhero    & 10 & 17 & 54 & 27 & 108 & 129 \\
        Financial    & 8  & 30 & 76 & 38 & 152 & 106 \\
        Student Club & 8  & 33 & 82 & 41 & 164 & 158 \\
        Card Games   & 6  & 47 & 106& 23 & 182 & 191 \\
        Formula 1    & 13 & 39 & 104& 52 & 208 & 174 \\
    \bottomrule
    \end{tabular}
    \label{tab:bird_synth}
\end{table}

\subsection{Synthetic data lake creation}
\label{sec:synth_data_lake_creation}
To simulate the redundancy, fragmentation, and quality degradation typical of enterprise data lakes, we developed a data generation pipeline that transforms the BIRD dataset into a noisy table repository via three noise-injection operations:

\smallskip
\noindent \textbf{Partitioned Tables.}
We used an LLM to identify logical partitioning keys (e.g., categorical regions or temporal buckets) and horizontally split base tables into multiple sub-tables. This forces the agent to compute unions across partitions to retrieve complete entities. For example, horizontally splitting a base \texttt{card} table along a categorical \texttt{type} key yields three sub-tables: \texttt{card\_type\_classic}, \texttt{card\_type\_junior}, and \texttt{card\_type\_gold}.

\smallskip
\noindent \textbf{Duplicate Tables.}
We duplicated tables to simulate different data lifecycle stages, creating production, staging, and testing variants. 
A \emph{production table} is an exact, clean copy of the original table representing the authoritative source. A \emph{staging table} duplicates the base table with injected noise, either by (1) appending a random 10\% sample of duplicate rows, or (2) replacing values with \texttt{NULL} at a 5\% probability rate. A \emph{testing table} contains a random 10\% row sample of the original table.
This setup forces the system to reason about data lineage to select the authoritative view, as we consider only the clean \texttt{production} tables to be correct.

\smallskip
\noindent \textbf{Low-quality Tables.}
We intentionally degraded data integrity by generating tables with broken foreign keys, injected random NULLs, injected duplicate rows, and missing entities.
We broke foreign keys by replacing 10\% of row references with \texttt{NULL} and another 5\% with an out-of-bounds index (e.g., \texttt{99999999}) for columns ending in \texttt{\_id}, simulating orphan records.
We injected NULLs by randomly masking 5\% of cell values across the table with \texttt{NULL}.
We injected duplicate data by randomly sampling 10\% of rows and appending them back to the table. 
We introduced incompleteness by extracting a 20\% random row sample from the table to generate subset tables.

To track this data generation pipeline, we recorded detailed lineage metadata for all derived tables. For partitioned tables, we stored a human-readable explanation of the partition rule (e.g., ``Split partition of seasons where year is 1970''), the original source table name, the partition column, and the split data type, following a systematic naming convention: \texttt{\{source\}\_\{column\}\_\{value\}}. For duplicate tables, in addition to appending lifecycle suffixes (\texttt{\_prod}, \texttt{\_stg}, and \texttt{\_test}), we added descriptions such as ``Clean PROD version'', ``Sample TEST version'', or ``Dirty STG version'', while preserving a reference to the base table. Finally, for low-quality tables, each variant retained its source table reference alongside a degradation description (e.g., ``Broke FK columns in \{column\}'', ``Injected random NULLs into \{table\}'', ``Duplicated rows to \{table\}'', or ``A subset of \{table\}'') and was labeled with a corresponding suffix (\texttt{\_broken\_fk}, \texttt{\_nulls}, \texttt{\_dups}, or \texttt{\_subset}). We incorporated this lineage information into the metadata description of each table, making it available to \name and all evaluated baselines.

\subsection{Baselines}

We compare \name against five baseline approaches:
\begin{itemize}
    \item \textbf{Vector Search}: This baseline serves as a non-agentic retrieval control. It embeds candidate metadata chunks into vectors, applies a 0.7 cosine-distance threshold to discard irrelevant results, and re-orders the remaining candidates with a semantic ranker. Because of its strong retrieval performance, \name also uses this vector-search engine with discrimination-oriented metadata as its \texttt{search()} tool.
    \item \textbf{Pneuma}~\citep{balaka2025pneuma}: Pneuma performs hybrid full-text and vector search over LLM-generated metadata, followed by an LLM-based judge that re-ranks the top-$K$ tables. We implement it with Gemini 3.5 Flash over the same documentation as \name.
    \item \textbf{Deterministic Workflow with LLM-powered Tool Calls}: This non-agentic baseline executes a fixed pipeline over the same metadata and tools as \name. First, it retrieves the top-10 candidate tables via \texttt{search()} and enriches them with attached and on-the-fly metadata by using an LLM to parameterize and invoke the specialized verification tools. Finally, a concluding LLM call selects the relevant subset of tables.
    \item \textbf{DS-Guru Dataset Selector}~\citep{lai2025kramabench}: DS-Guru selects tables via a single-shot in-context prompt by concatenating the schemas and sample rows of up to 100 candidate tables into one LLM call. While effective for small catalogs where all metadata fits within the prompt, this approach fails to scale to larger enterprise data lakes due to context-window truncation and attention dilution. Consequently, in high-cardinality environments, DS-Guru achieves low accuracy.
    \item \textbf{Agentic SQL Dataset Selector}: An agentic multi-turn baseline operating directly over full SQL schema representations. It uses an interactive ReAct~\citep{yao2022react} loop to query database catalogs (via \texttt{INFORMATION\_SCHEMA}) and sample table contents dynamically via raw SQL queries. Unlike \name, which prunes candidate tables using pre-indexed discrimination-oriented vector search and attached metadata, Agentic SQL lacks an initial semantic retrieval stage and relies entirely on iterative SQL exploration over the raw catalog.
\end{itemize}

\noindent We also evaluate four \name configurations by comparing the full method against three ablation variants, which serve as strong agentic retrieval and reasoning baselines:
\begin{itemize}
    \item \textbf{MR-\footnotesize{Search}}: Uses discrimination-oriented metadata within the agentic search workflow. The agent may perform multiple search iterations, but consumes responses only from the search engine.
    \item \textbf{MR-\footnotesize{Search+Attached}}: Incorporates attached table summaries in the prompt without invoking specialized tools for on-the-fly metadata.
    \item \textbf{MR-{\footnotesize Search+On-the-fly}}: Omits attached table metadata from the LLM prompt, testing the system's ability to rely on on-the-fly metadata alongside semantic search for reasoning.
    \item \textbf{Full MR~(MR-\footnotesize{Search+Attached+On-the-fly})}: The complete architecture that autonomously combines semantic search and fetches attached summaries and on-the-fly metadata.
\end{itemize}

Unless noted otherwise, all LLM-driven baselines and \name variants are evaluated using Gemini 3.8 Flash as the backbone model. 
All evaluated methods access the same enriched metadata (including lineage information and content summaries) and use the \texttt{text-embedding-005} model for metadata encoding.
We implement the agentic variants using the Agent Development Kit framework\footnote{ADK: \url{https://google.github.io/adk-docs/}} and the Google Antigravity harness platform\footnote{Google Antigravity: \url{https://antigravity.google/}}.

\subsection{Implementations of \name}
\label{sec:impl_setup}
\label{sec:implementations}
Because the landscape of data agents is evolving rapidly, an open question is whether metadata reasoning should operate as a standalone agent or as part of a general-purpose agent harness. We therefore evaluate two implementations of \name: a standalone agent and a plugin for a general agent harness that performs data-related tasks.
The two implementations share the same semantic search engine, attached and on-the-fly metadata, and selection criteria, but differ in which agent orchestrates the reasoning loop. 

\smallskip
\noindent\textbf{Dedicated Agent.}
The Dedicated Agent encapsulates the discovery workflow within a standalone agent that owns its prompt, state, and iterative reasoning controller. Other agents invoke it either as a single tool or as a subagent, passing an analytical task and receiving the selected tables and justification in one call.

\smallskip
\noindent\textbf{Harness Plugin.}
In the Harness Plugin, the host agent of a general-purpose agent harness (such as Google Antigravity, Codex, or Claude Code) performs the metadata reasoning itself. The plugin provides three modular components: (1) an \emph{MCP server} that exposes each catalog operation as a separate tool, including semantic search, metadata retrieval, the specialized tools for fetching on-the-fly metadata, and recommendation recording, (2) \emph{agent skills} that guide the host agent through the overall selection procedure, coverage, data quality, join paths, value verification, and interleaved search and verification, and (3) \emph{behavioral rules} that enforce four invariants: selecting only tables observed in tool outputs, formulating search queries in descriptive natural language, pruning candidates only with cited evidence, and returning a structured justification.

\subsection{Evaluation methods and metrics}
\subsubsection{Evaluation with ground-truth table sets}
\label{sec:eval_gt}
Standard rank-based metrics (e.g., Hit@$K$) are ill-suited for agentic systems that output variable-length sets. We therefore evaluate selection accuracy between the reference table set $T_{\mathrm{ref}}$ and the predicted table set $T_{\mathrm{pred}}$ using a set-aware \textbf{F1 score}, the harmonic mean of precision and recall. Specifically, recall measures the sufficiency (coverage) of the selected data sources, and precision measures their minimality (conciseness). To ensure a rigorous comparison with ranking-based baselines such as vector search, we evaluate their performance across various retrieval thresholds $K$ and report the highest achieved F1 score. The metrics are defined as follows: 
\begin{itemize}
\item \textbf{Recall}: $\frac{|T_{\mathrm{ref}} \cap T_{\mathrm{pred}}|}{|T_{\mathrm{ref}}|}$ (measures coverage), 
\item \textbf{Precision}: $\frac{|T_{\mathrm{ref}} \cap T_{\mathrm{pred}}|}{|T_{\mathrm{pred}}|}$ (measures conciseness), 
\item \textbf{F1 score}: $2 \times \frac{\text{Precision} \times \text{Recall}}{\text{Precision} + \text{Recall}}$.
\end{itemize}
We also report the average number of reasoning steps per task (\textbf{Avg.\ Steps}) to measure reasoning overhead.

\subsubsection{Evaluation without ground-truth table sets}

Evaluating table selection accuracy on the synthetic BIRD data lake is challenging because derived tables lack direct ground-truth reference labels. However, by leveraging our lineage tracking, which maps every derived table back to its clean base ancestor, we use the dataset's existing gold SQL queries to systematically evaluate table selection performance without manual re-labeling. Our evaluation procedure follows three steps:

\noindent \textbf{1. Parse Gold Query}: We extract all base table references and filter constraints (e.g., \texttt{WHERE} clauses) from the gold SQL query.

\noindent \textbf{2. Trace Lineage}: For each selected table, we consult the lineage map to identify its clean base ancestor and the specific transformations used to generate it (e.g., horizontal splits or noise injection).

\noindent \textbf{3. Verify Correctness}: A selected table is scored as correct only if it satisfies three conditions: (1) its associated base ancestor is required by the gold SQL query, (2) it excludes undesirable quality markers and noise suffixes (\texttt{\_stg}, \texttt{\_test}, \texttt{\_subset}, \texttt{\_dups}, \texttt{\_broken\_fk}, or \texttt{\_nulls}), and (3) for horizontally split partitions, its partition values match the filter constraints in the gold SQL query (e.g., \texttt{district\_A3\_east\_bohemia\_prod} is correct if the gold SQL query contains \texttt{WHERE A3 = "East Bohemia"} on the \texttt{District} table).

Based on the verified tables in the selected set, we calculate recall, precision, and F1 score for each query.

%% file: sections/5_experimental_evaluation.tex
\section{Experimental Evaluation}
\label{sec:experiments}
We evaluate \name across the real-world KramaBench environment and the synthetically scaled BIRD dataset. We structure our evaluation around: (1) end-to-end selection accuracy, (2) the ablation of attached and on-the-fly metadata, (3) robustness against structural noise, (4) metadata representations for semantic search, (5) reasoning cost and efficiency, and (6) an analysis of the remaining selection errors.

\subsection{End-to-end data selection performance}
We first evaluate whether \name accurately identifies the required combination of data sources.

\begin{table}[tb]
\centering
\caption{Average recall (Rec.), precision (Prec.), F1 score, and number of steps on KramaBench across methods.}
\label{tab:average_performance}
    \resizebox{\linewidth}{!}{
    \begin{tabular}{l |rrr|r}
    \toprule
    \textbf{Method} & \textbf{Rec. (\%)} & \textbf{Prec. (\%)} & \textbf{F1 (\%)} & \textbf{Avg. Steps} \\
    \midrule
    \textbf{Vector Search} & - & - & 50.77 & - \\
    \textbf{Pneuma} & - & - & 45.12 & - \\
    \textbf{Deterministic} & 67.32\tiny{$\pm$0.47} & 73.88\tiny{$\pm$0.68} & 66.25\tiny{$\pm$0.50} & 7.05\tiny{$\pm$0.04} \\
    \textbf{DS-Guru} & 76.28\tiny{$\pm$1.17} & 73.74\tiny{$\pm$0.90} & 73.33\tiny{$\pm$0.95} & 1.00\tiny{$\pm$0.00} \\
    \textbf{Agentic SQL} & 84.01\tiny{$\pm$1.83} & 80.27\tiny{$\pm$3.26} & 77.90\tiny{$\pm$2.17} & 10.20\tiny{$\pm$0.25} \\
    \midrule
    \multicolumn{5}{l}{\textit{Dedicated Agent}} \\
    \textbf{MR-\footnotesize{Search}} & 78.60\tiny{$\pm$2.52} & 85.47\tiny{$\pm$1.31} & 79.60\tiny{$\pm$1.53} & 11.44\tiny{$\pm$0.13} \\
    \textbf{MR-\footnotesize{Search+Attached}} & 82.85\tiny{$\pm$0.97} & 87.98\tiny{$\pm$1.15} & 83.33\tiny{$\pm$1.01} & 9.37\tiny{$\pm$0.13} \\
    \textbf{MR-\footnotesize{Search+On-the-fly}} & 78.88\tiny{$\pm$1.73} & 83.83\tiny{$\pm$0.59} & 78.68\tiny{$\pm$1.18} & 18.28\tiny{$\pm$0.26} \\
    \textbf{Full MR} & 87.25\tiny{$\pm$0.46} & 90.78\tiny{$\pm$0.22} & 87.09\tiny{$\pm$0.40} & 10.49\tiny{$\pm$0.25} \\
    \midrule
    \multicolumn{5}{l}{\textit{Harness Plugin}} \\
    \textbf{MR-\footnotesize{Search}} & 82.86\tiny{$\pm$0.25} & 86.24\tiny{$\pm$0.15} & 82.27\tiny{$\pm$0.18} & 6.95\tiny{$\pm$0.12} \\
    \textbf{MR-\footnotesize{Search+Attached}} & 86.45\tiny{$\pm$1.54} & 88.74\tiny{$\pm$1.47} & 85.47\tiny{$\pm$1.38} & 6.82\tiny{$\pm$0.16} \\
    \textbf{MR-\footnotesize{Search+On-the-fly}} & 84.86\tiny{$\pm$1.30} & 87.10\tiny{$\pm$1.50} & 83.54\tiny{$\pm$1.41} & 10.59\tiny{$\pm$0.22} \\
    \textbf{Full MR} & \textbf{89.44}\tiny{$\pm$0.37} & \textbf{91.73}\tiny{$\pm$0.54} & \textbf{88.63}\tiny{$\pm$0.14} & 8.00\tiny{$\pm$0.08} \\
    \bottomrule
    \end{tabular}
}
\end{table}

\begin{table*}[tb]
\centering
\caption{F1 score (\%) comparison across six domains. The number in parentheses after each domain name indicates the number of candidate tables. The F1 scores for the vector search baseline represent the best F1 scores achieved across various choices of top-$K$ results.}
\label{tab:f1_breakdown}
\resizebox{\linewidth}{!}{
    \begin{tabular}{l | ccc | cccc | cccc | cc}
    \toprule
    \multirow{3}{*}{\textbf{Domain}} & \multirow{3}{*}{\begin{tabular}[c]{@{}c@{}}\textbf{Vector}\\\textbf{Search}\end{tabular}} & \multirow{3}{*}{\begin{tabular}[c]{@{}c@{}}\textbf{DS-}\\\textbf{Guru}\end{tabular}} & \multirow{3}{*}{\begin{tabular}[c]{@{}c@{}}\textbf{Agentic}\\\textbf{SQL}\end{tabular}} & \multicolumn{4}{c|}{\textbf{MR (Dedicated, Gemini 3.8 Flash)}} & \multicolumn{4}{c|}{\textbf{MR (Plugin, Gemini 3.8 Flash)}} & \multicolumn{2}{c}{\textbf{Full MR (Sonnet 5)}} \\
    \cmidrule(lr){5-8} \cmidrule(lr){9-12} \cmidrule(lr){13-14}
    & & & & \footnotesize{Search} & \footnotesize{+Attached} & \footnotesize{+On-the-fly} & \footnotesize{Full} & \footnotesize{Search} & \footnotesize{+Attached} & \footnotesize{+On-the-fly} & \footnotesize{Full} & \footnotesize{Dedicated} & \footnotesize{Plugin} \\
    \midrule
    Astronomy\tiny{(1498)}   & 32.80 & 12.50\tiny{$\pm$0.00} & 64.54\tiny{$\pm$7.84} & 68.25\tiny{$\pm$4.26} & 76.06\tiny{$\pm$1.88} & 71.90\tiny{$\pm$4.41} & 79.33\tiny{$\pm$2.89} & 73.96\tiny{$\pm$0.15} & 78.31\tiny{$\pm$3.16} & 74.48\tiny{$\pm$1.47} & 85.90\tiny{$\pm$4.68} & 74.43\tiny{$\pm$5.07} & \textbf{86.45}\tiny{$\pm$2.18} \\
    Legal\tiny{(134)}       & 38.45 & 59.86\tiny{$\pm$1.39} & 58.49\tiny{$\pm$3.63} & 64.97\tiny{$\pm$0.64} & 64.42\tiny{$\pm$0.64} & 61.83\tiny{$\pm$0.32} & \textbf{72.91}\tiny{$\pm$0.32} & 67.20\tiny{$\pm$0.64} & 66.17\tiny{$\pm$0.90} & 63.49\tiny{$\pm$1.28} & 72.26\tiny{$\pm$1.40} & 64.65\tiny{$\pm$0.91} & 70.79\tiny{$\pm$1.73} \\
    Environment\tiny{(36)} & 45.56 & 94.83\tiny{$\pm$0.42} & 89.58\tiny{$\pm$2.40} & 85.19\tiny{$\pm$2.65} & 86.58\tiny{$\pm$1.28} & 80.79\tiny{$\pm$5.24} & 93.57\tiny{$\pm$2.42} & 91.04\tiny{$\pm$0.00} & 89.25\tiny{$\pm$2.75} & 90.54\tiny{$\pm$4.16} & 93.88\tiny{$\pm$2.42} & 86.73\tiny{$\pm$1.38} & \textbf{96.26}\tiny{$\pm$0.01} \\
    Biomedical\tiny{(29)}  & 50.37 & \textbf{90.85}\tiny{$\pm$5.22} & 75.59\tiny{$\pm$4.12} & 80.49\tiny{$\pm$2.14} & 83.46\tiny{$\pm$4.28} & 77.78\tiny{$\pm$2.57} & 85.93\tiny{$\pm$0.00} & 80.49\tiny{$\pm$2.14} & 88.36\tiny{$\pm$7.33} & 87.41\tiny{$\pm$5.13} & 88.15\tiny{$\pm$3.85} & 83.24\tiny{$\pm$6.58} & 79.26\tiny{$\pm$3.85} \\
    Wildfire\tiny{(21)}    & 58.25 & 86.95\tiny{$\pm$4.32} & 84.99\tiny{$\pm$6.35} & 83.33\tiny{$\pm$0.82} & 95.03\tiny{$\pm$2.00} & 82.54\tiny{$\pm$2.44} & 96.35\tiny{$\pm$1.37} & 83.70\tiny{$\pm$2.11} & 96.30\tiny{$\pm$1.47} & 89.05\tiny{$\pm$1.67} & \textbf{97.14}\tiny{$\pm$0.00} & 93.44\tiny{$\pm$0.92} & 95.56\tiny{$\pm$0.00} \\
    Archaeology\tiny{(5)}  & 79.17 & 95.00\tiny{$\pm$1.24} & 94.22\tiny{$\pm$4.47} & 95.37\tiny{$\pm$1.60} & 94.44\tiny{$\pm$0.00} & 97.22\tiny{$\pm$0.00} & 94.44\tiny{$\pm$0.00} & \textbf{97.22}\tiny{$\pm$0.00} & 94.44\tiny{$\pm$0.00} & 96.30\tiny{$\pm$1.60} & 94.44\tiny{$\pm$0.00} & 93.52\tiny{$\pm$1.60} & \textbf{97.22}\tiny{$\pm$0.00} \\
    \midrule
    Avg. F1  & 50.77 & 73.33\tiny{$\pm$0.95} & 77.90\tiny{$\pm$2.17} & 79.60\tiny{$\pm$1.53} & 83.33\tiny{$\pm$1.01} & 78.68\tiny{$\pm$1.18} & 87.09\tiny{$\pm$0.40} & 82.27\tiny{$\pm$0.18} & 85.47\tiny{$\pm$1.38} & 83.54\tiny{$\pm$1.41} & \textbf{88.63}\tiny{$\pm$0.14} & 82.67\tiny{$\pm$1.34} & 87.59\tiny{$\pm$0.40} \\
    \bottomrule
    \end{tabular}
}
\end{table*}

\subsubsection{Results on KramaBench}
Table~\ref{tab:average_performance} lists the average accuracy scores on KramaBench with Gemini 3.8 Flash as the backbone model. The metrics are calculated per domain and averaged across the six domains. All LLM-powered methods are run three times, and we report the mean and standard deviation across runs.

As shown in Table~\ref{tab:average_performance}, Full MR achieves the highest overall F1 score in both the Dedicated Agent and Harness Plugin implementations. Notably, every ablated variant of \name in both implementations also outperforms all five baselines. The retrieval methods perform poorly because ranking individual tables by semantic similarity alone cannot verify multi-hop join paths or filter out tables that lack required attributes or entities. The Deterministic Workflow baseline improves over retrieval by executing a fixed sequence of tool calls, but it falls short of \name because it cannot dynamically reformulate searches when initial candidates are incomplete.

The domain breakdowns in Table~\ref{tab:f1_breakdown} reveal that \name's adaptive reasoning outperforms LLM baselines across diverse catalog structures. DS-Guru, which relies on a single LLM call, slightly edges out \name on small catalogs, e.g., Biomedical, because the entire domain schema fits within the context window, allowing the model to compare all tables simultaneously without retrieval loss. However, in large repositories such as Astronomy, context window limits force aggressive candidate truncation prior to reasoning, causing a severe drop in accuracy and confirming that single-call prompt stuffing cannot scale to enterprise data lakes. Finally, Agentic SQL is the strongest baseline because it explores the catalog interactively. However, while it achieves competitive recall, its precision lags behind \name because ad-hoc SQL probing over raw schemas lacks precomputed discrimination-oriented summaries and specialized tools that fetch on-the-fly metadata to verify candidate tables, leading the agent to retain unverified candidate tables.

To evaluate model generality, Table~\ref{tab:f1_breakdown} also includes Full \name evaluated with Claude Sonnet 5 as the backbone model across both implementations. Claude Sonnet 5 demonstrates robust domain performance, achieving an average F1 score of 82.67\% for the Dedicated Agent and 87.59\% for the Harness Plugin, outperforming all LLM baselines and reaching top performance in Environment at 96.26\% and Astronomy at 86.45\%.

\smallskip
\noindent \textbf{Minimality of Selected Table Sets.} Across KramaBench, \name selects nearly the same number of tables as the ground truth, rarely returning unnecessary tables. In four of the six domains (Archaeology, Biomedical, Environment, and Wildfire), the average size of the selected set closely matches the ground truth, reflecting \name's high precision (up to 91.73\% in Table~\ref{tab:average_performance}) in pruning redundant candidates. A gap in average set size appears only in Astronomy and Legal, where a few tasks require combining dozens to over a hundred partitioned tables (e.g., metropolitan reports in Legal or daily orbital files in Astronomy) and \name retrieves only a subset of those partitions within its default step budget. Allowing a few additional reasoning steps closes this gap in Astronomy by letting the agent enumerate the full partition family, whereas recovering over a hundred partitions in Legal remains challenging under a fixed step limit.

\begin{table}[bt]
\centering
\caption{Data selection accuracy (\%) of \name (MR) and top-10 vector search on the five synthetic BIRD datasets.}
\label{tb:synth_breakdown_res}
\resizebox{\linewidth}{!}{
    \begin{tabular}{l l c rrr}
    \toprule
    \textbf{Dataset} & \textbf{Version} & \textbf{Method} & \textbf{Precision} & \textbf{Recall} & \textbf{F1 score} \\ \midrule
    \multirow{3}{*}{Superhero} & Base (10) & MR & 100.0 & 98.5 & 99.0 \\ \cline{2-6}
    & \multirow{2}{*}{Messy (108)} & MR & 99.6 & 97.7 & \textbf{98.3} \\
    & \multicolumn{1}{c}{} & Top-10 & 25.1 & 66.8 & 35.2 \\ \midrule
    
    \multirow{3}{*}{Financial} & Base (8) & MR & 93.6 & 89.8 & 90.1 \\ \cline{2-6}
    & \multirow{2}{*}{Messy (152)} & MR & 92.4 & 87.9 & \textbf{88.4} \\
    & \multicolumn{1}{c}{} & Top-10 & 29.7 & 58.1 & 37.2 \\ \midrule

    \multirow{3}{*}{Student Club} & Base (8) & MR & 100.0 & 99.3 & 99.5 \\ \cline{2-6}
    & \multirow{2}{*}{Messy (164)} & MR & 94.9 & 93.9 & \textbf{94.3} \\
    & \multicolumn{1}{c}{} & Top-10 & 17.7 & 56.3 & 25.7 \\ \midrule

    \multirow{3}{*}{Card Games} & Base (6) & MR & 97.9 & 97.0 & 97.1 \\ \cline{2-6}
    & \multirow{2}{*}{Messy (182)} & MR & 96.0 & 95.1 & \textbf{95.3} \\
    & \multicolumn{1}{c}{} & Top-10 & 14.7 & 51.1 & 21.4 \\ \midrule
    
    \multirow{3}{*}{Formula 1} & Base (13) & MR & 93.9 & 93.8 & 93.5 \\ \cline{2-6}
    & \multirow{2}{*}{Messy (208)} & MR & 91.4 & 91.4 & \textbf{91.0} \\
    & \multicolumn{1}{c}{} & Top-10 & 20.4 & 66.9 & 30.3 \\ \midrule \midrule

    \multirow{3}{*}{\textbf{Average}} & Base & MR & 97.1 & 95.7 & 95.8 \\ \cline{2-6}
    & \multirow{2}{*}{Messy} & MR & 94.9 & 93.2 & \textbf{93.5} \\
    & \multicolumn{1}{c}{} & Top-10 & 21.5 & 59.8 & 30.0 \\ \bottomrule
    \end{tabular}
}
\end{table}

\subsubsection{Results on the synthetic dataset}
Table~\ref{tb:synth_breakdown_res} evaluates end-to-end selection accuracy of the Harness Plugin implementation on the synthetically scaled BIRD data lake with Gemini 3.8 Flash as the backbone model.
Across all five datasets, \name remains accurate even as horizontal partitioning and noise injection expand the number of candidate tables by up to 30$\times$: its average F1 score drops by only 2.3 percentage points, from 95.8\% on the clean base schemas to 93.5\% on the messy data lake. In contrast, the top-10 vector search baseline reaches only 30.0\% average F1 on the messy data lake, with both low precision (21.5\%) and low recall (59.8\%). Its precision suffers because it returns all 10 retrieved tables without filtering out noisy copies, while its recall drops because near-duplicate noisy tables crowd the top-10 results and push required tables outside the cutoff.

\subsection{Ablation of the attached and on-the-fly metadata} 
We conduct an ablation study to evaluate the effectiveness of the attached and on-the-fly metadata (Tables~\ref{tab:average_performance} and~\ref{tab:f1_breakdown}).

\smallskip
\noindent \textbf{Performance Boost from Attached Metadata.} Incorporating attached table content summaries (\emph{MR-{\footnotesize Search+Attached}}) provides the largest standalone improvement in selection accuracy while simultaneously reducing or maintaining the number of reasoning steps across both implementations compared to using semantic search alone (\emph{MR-{\footnotesize Search}}). Attached summaries act as an immediate warm start that helps the agent disambiguate relevant tables and prune obvious false positives in a single reasoning pass without issuing additional exploratory queries.

\smallskip
\noindent \textbf{On-the-fly Metadata Trade-off.} An interesting pattern in Table~\ref{tab:average_performance} is that enabling specialized tools that fetch on-the-fly metadata without attached summaries (\emph{MR-{\footnotesize Search+On-the-fly}}) sharply inflates the reasoning step count while providing little to no accuracy gain over search alone. Deprived of prior statistical context from attached profiles, the agent cannot form targeted verification hypotheses and instead resorts to costly trial-and-error tool probing across candidate columns. Conversely, when on-the-fly tools are paired with attached metadata (\emph{Full MR}), the two components complement each other, achieving the best overall precision, recall, and F1 score across both implementations with only minimal additional steps over \emph{MR-{\footnotesize Search+Attached}}. Attached profiles narrow the ambiguity to specific candidate columns or table groups, allowing the agent to invoke tools selectively. This complementary behavior is especially pronounced in domains whose answers span partitioned or redundant table families (Astronomy, Legal, and Environment in Table~\ref{tab:f1_breakdown}), where \texttt{find\_related\_tables()} enumerates sibling shards of a table family in a single call and \texttt{joinability\_check()} measures key overlap to eliminate redundant summary tables.

\subsection{Robustness against data messiness}

\begin{figure}[tb]
    \centering
    \includegraphics[width=\linewidth]{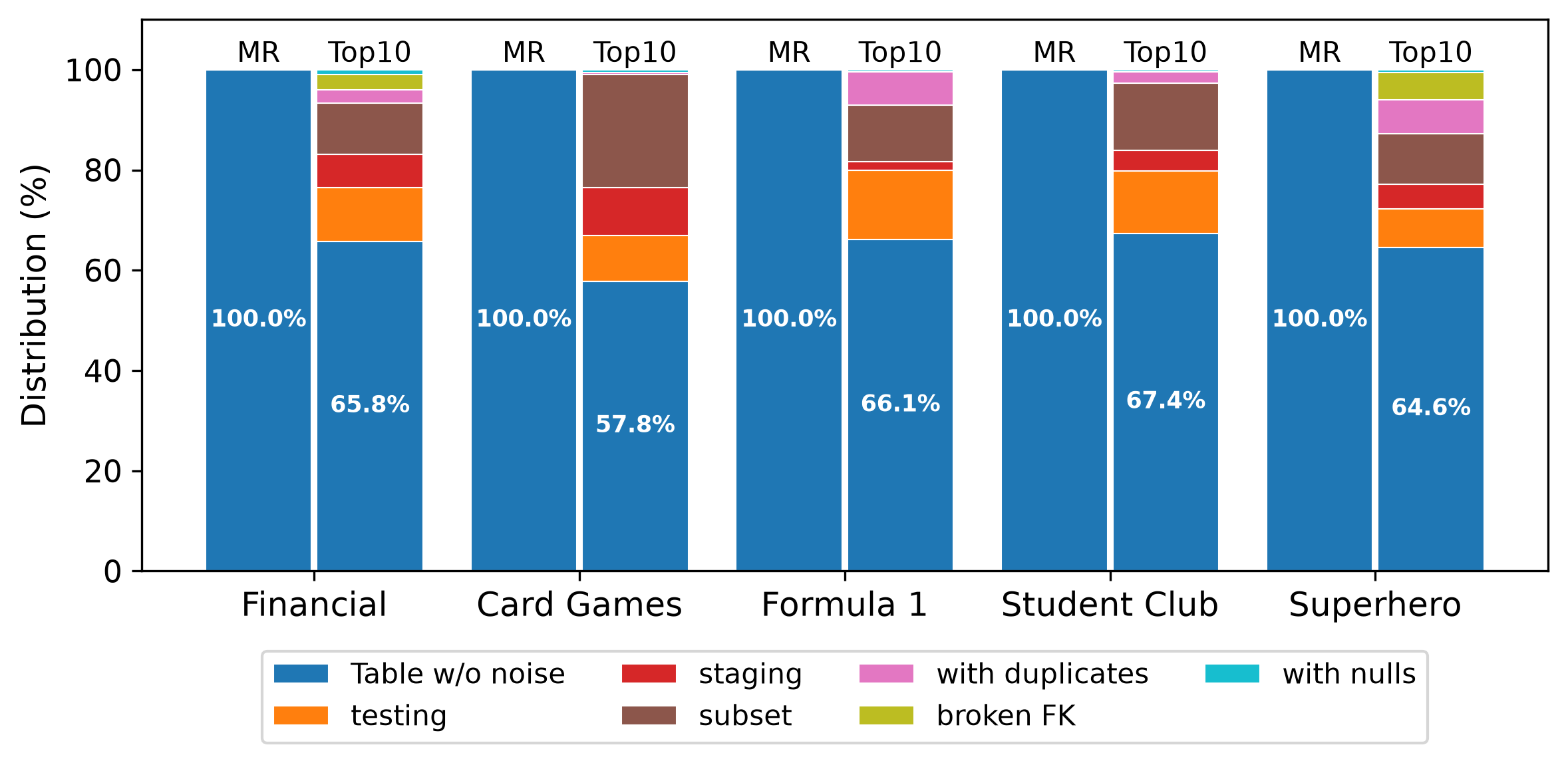}
    \vspace{-4pt}
    \caption{Distribution of selected table types from \name (MR) and top-10 search results in the synthetic messy data lake (\%). \name successfully selected 100.0\% of the tables without injected noise.}
    \label{fig:syn_selected_table_ratio}
    \vspace{-8pt}
\end{figure}

To explain why ranking-based retrieval engines fail in complex enterprise environments, we analyzed the distribution of table types selected by both systems across our noisy synthetic data lake. As shown in Figure~\ref{fig:syn_selected_table_ratio}, \name exclusively selects clean, authoritative production and base tables (100.0\% clean tables) while completely avoiding staging, testing, incomplete subset, or corrupted tables (0.0\% noisy variants) across every domain. In contrast, a substantial portion of the tables retrieved by the vector search baseline consists of noisy variants, predominantly incomplete subset snapshots and test splits.

This performance gap highlights a fundamental limitation of text embeddings: because staging copies, subset snapshots, and test splits preserve the same schemas and text as authoritative production tables, embedding geometry cannot separate clean tables from corrupted replicas. \name overcomes this conflation through a two-stage defense. During retrieval, discrimination-oriented indexing separates similar tables by contrasting their relative differences. During reasoning, \name inspects attached lineage metadata and invokes specialized tools to fetch on-the-fly metadata (\texttt{column\_profiler()} and \texttt{data\_finder()}) to verify row counts, NULL distributions, and temporal coverage, filtering out degraded replicas that look identical at the schema level.

\subsection{Effectiveness of discrimination-oriented search}
\label{app:RQ5_metadata_search}
We evaluate the initial retrieval stage of our pipeline in isolation to validate our embedding representation strategy. When building the vector index, embedding raw schemas or standalone table summaries often causes semantically related tables to cluster together in the vector space, degrading retrieval recall when a query spans multiple tables or partitions. To ensure a fair comparison, we encode four metadata representations with the same text embedding model (\texttt{text-embedding-005}) and query them with the same vector search engine:
\begin{itemize}
    \item \textbf{Schema only}: Embeds the raw table schema, consisting of the table name, column names and data types, and three randomly sampled rows for basic value grounding.
    \item \textbf{Table content-summarization}: Embeds a single-table LLM summary generated from the physical schema and statistical column profiles (value ranges, top-$K$ categories, and null ratios), following the enrichment procedure of Section~\ref{sec:tier2_metadata}.
    \item \textbf{AutoDDG}~\citep{zhang2025autoddg}: Embeds descriptions produced by an automated multi-stage profiling pipeline that combines topic generation, semantic profiling across temporal, spatial, entity, and domain concepts, unformatted description synthesis, and structured description expansion for each table independently.
    \item \textbf{Discrimination-oriented metadata}: Embeds descriptions produced by the two-stage group-aware pipeline of Section~\ref{sec:method_discriminative_description}.
\end{itemize}

\begin{table}[t]
\centering
\small
\caption{Average Recall@K (\%) across the six KramaBench domains for different metadata representations. ``Rec@K'' denotes the percentage of ground-truth tables recovered within the top-$K$ retrieved results (for Archaeology, which contains 5 tables in total, Rec@10 equals Rec@5 at 100.0\%).}
\label{tab:recall_scores_search}
\resizebox{\linewidth}{!}{
    \begin{tabular}{lccc}
    \toprule
    \textbf{Metadata Representation} & \textbf{Rec@1 (\%)} & \textbf{Rec@5 (\%)} & \textbf{Rec@10 (\%)} \\
    \midrule
    Schema Only           & 30.56 & 57.70 & 68.32 \\
    Table Content-Summarization      & \textbf{40.58} & 62.07 & 72.52 \\
    AutoDDG & 36.24 & 64.08 & 73.70 \\
    Discrimination-Oriented          & 40.53 & \textbf{70.71} & \textbf{75.47} \\
    \bottomrule
    \end{tabular}
}
\end{table}

Table~\ref{tab:recall_scores_search} compares retrieval recall across cutoffs for the four metadata representations.
Discrimination-oriented metadata achieves the strongest overall retrieval performance: it matches the top single-table method at Recall@1 while leading all methods at Recall@5, which is 6.63 percentage points over AutoDDG, and Recall@10, which is 1.77 percentage points over the best baseline.
Across all cutoffs, the schema-only baseline performs worst because abbreviated column headers and a few sample rows lack the semantic context needed to bridge the vocabulary gap with natural-language queries.
Enriching tables with LLM-generated descriptions improves retrieval over raw schemas, with an instructive trade-off between the two single-table baselines. Table content-summarization performs best on Recall@1, whereas AutoDDG overtakes it at Recall@5 and Recall@10 by expanding descriptions with broader thematic and conceptual profiles. However, because both approaches summarize each table in isolation, sibling partitions and near-duplicate tables receive nearly identical embeddings, causing them to cluster tightly in vector space and limiting recall gains as $K$ increases.

\subsection{Cost and efficiency analysis}
\label{sec:cost_efficiency}

To quantify the operational overhead of deploying \name, we analyze its reasoning steps and token consumption.

\smallskip
\noindent \textbf{Step Count as a Hardware-Agnostic Metric.} Because wall-clock latency fluctuates with serving infrastructure, network conditions, API rate limits, and candidate retrieval counts, we report the average number of reasoning steps per query in Table~\ref{tab:average_performance} as a hardware- and platform-agnostic efficiency metric. Using attached metadata as a warm start allows Full MR to converge in substantially fewer steps than tool orchestration without attached metadata (\emph{MR-{\footnotesize Search+On-the-fly}}) in both the Dedicated Agent (10.49 vs.\ 18.28 steps) and the Harness Plugin (8.00 vs.\ 10.59 LLM calls), avoiding the multi-turn overhead of unguided schema and value probing.

\smallskip
\noindent \textbf{Token Cost Accounting.} During semantic retrieval, our discrimi\-nation-oriented descriptions average 175 tokens per table. During reasoning, attached metadata profiles average 1,308 tokens per table across KramaBench domains. Passing these compact statistical summaries into the prompt instead of raw database rows prevents context saturation and keeps prompt costs low.

\smallskip
\noindent \textbf{Operational Cost Trade-offs.} Practical deployment cost scales with total token traffic across turns. With the Gemini 3.8 Flash Harness Plugin, Full MR averages 8.00 LLM calls per task and consumes a cumulative $\sim$178.7k prompt tokens ($\sim$22.3k per call) and $\sim$1.3k completion tokens per task.

%% file: sections/5_failure_analysis.tex
\subsection{Error analysis}
\label{sec:failure_analysis}

We analyze the selection errors of the Harness Plugin Full MR with Gemini 3.8 Flash, the configuration of Table~\ref{tab:average_performance}, in one run over the 104 KramaBench tasks. Twenty-five tasks have an F1 score below 100\%, and they account for the 13.0-point difference between the task-mean F1 score of 87.0\% and 100\%. We refer to this difference as the F1 gap and manually assign each of the 25 tasks to one of five error categories.

The category with the largest share of the F1 gap is \textbf{runtime-dependent selection}, which comprises 5 tasks and 31\% of the F1 gap. In these tasks, the correct member of a table collection is determined by a value that an upstream query computes, for example the three-day shard that contains the peak satellite density or the annual water-testing table of the wettest summer. Since \name does not execute queries, a single pass either returns the entire collection, for example 22 shards when 2 are required, or selects one member without evidence. Section~\ref{sec:integration} resolves this category by letting \name request the value from the host agent.

The remaining four categories are as follows. \textbf{Consolidated table versus partition family} comprises 4 tasks and 30\% of the F1 gap. The Legal catalog stores metropolitan-area reports both as one national table and as 52 per-state partitions at the same grain; \name selects the national table, whereas the reference solution unions the partitions. \textbf{Defensible alternative set} comprises 6 tasks and 21\%. \name selects a table set that differs from the reference set but answers the question, for example a single sheet after \texttt{data\_finder()} verified that the queried peptide occurs in only one of three candidate sheets. \textbf{Coverage verification miss} comprises 3 tasks and 11\%. \name omits a table that a required attribute depends on, such as the lookup table that maps historical actors to modern countries. \textbf{Over-inclusion} comprises 7 tasks and 7\%. \name returns every reference table and one unnecessary lookup or denominator table, so the mean F1 score of these tasks remains 86.0\%. Across the five categories, 10 of the 25 tasks reached the step budget of 12 LLM calls, compared with 20\% of all tasks.

%% file: sections/6_agent_integration.tex
\section{Interleaving Planning and Execution}
\label{sec:integration}

In the previous sections, \name operated in isolation: it selected tables in a single pass based purely on catalog metadata, without interacting with other agents or executing downstream analytical code. In an end-to-end analytics workflow, however, \name works alongside a \emph{host agent} that writes and executes SQL or Python code to answer the task, and \name passes the host agent a \emph{handoff} consisting of the selected tables and a selection justification. Integrating the two agents raises a central question: \emph{should \name run once before code execution, or should the two agents exchange intermediate results during execution?} Running \name once upfront fails when selecting subsequent tables depends on intermediate query results, the error category that accounts for the largest share of the F1 gap. Decomposing the question into sub-queries ahead of time does not resolve this limitation because the values that identify downstream tables do not exist in the catalog metadata and materialize only after the upstream query executes.

\subsection{Runtime value binding}
\label{sec:runtime_binding}

We define a \emph{runtime binding} as an intermediate result of code execution, such as a computed value, an entity list, or an observed data format, that is needed to select subsequent tables. A task is \emph{runtime-dependent} if selecting its tables requires at least one runtime binding. A runtime binding typically identifies one or more members of a \emph{table collection}, which can be a family of tables that share a schema and partition a dataset by an attribute such as year, or a group of joinable tables used together. For example, to compute the water-quality exceedance rate at a beach during the wettest summer on record~\citep{lai2025kramabench}, the host agent first queries a precipitation table to find the wettest year, and \name then selects that year's table from a collection of annual water-testing tables. Similarly, querying an initial table can yield a list of identifiers that determines which partition tables to join next, or reveal the exact format needed to choose a compatible table. Without the runtime binding, \name returns the entire collection, which includes the required table alongside many unneeded ones. Given the runtime binding, \name selects only the tables that code execution requires. In this sense, a table collection resembles a data source with a binding pattern~\citep{rajaraman1995answering}: the right member can be accessed only after the required input value is known.

\subsection{Integration strategies}
\label{sec:integration_architectures}

To answer the question raised above, we compare three integration strategies that differ in whether runtime bindings flow back from the host agent to \name during execution and, if so, which agent initiates the exchange. S1 is a single-pass baseline without any exchange, whereas S2 and S3 allow exchanges initiated by the host agent and by \name, respectively. In all three strategies, \name and the host agent run in separate contexts, and the host agent can query the database:
\begin{itemize}[leftmargin=*]
    \item \textbf{S1: Single-pass.} \name runs once before code execution. When a table depends on a runtime binding, \name returns the entire collection, leaving the host agent to resolve the binding on its own.
    \item \textbf{S2: Host-agent-driven.} \name first returns only the tables needed to compute the runtime binding and defers the dependent collection. After computing the binding, the host agent may re-invoke \name to resolve the deferred collection.
    \item \textbf{S3: Discovery-driven.} When a table depends on a runtime binding, \name pauses and sends the host agent an execution request specifying the binding and the tables needed to compute it. Once the host agent returns the binding, \name resumes selection.
\end{itemize}
S2 and S3 share the same prompt instructions for identifying runtime bindings, so any difference between them isolates the effect of which agent initiates the exchange.

\subsection{Evaluation}
\label{sec:integration_experiments}

\noindent\textbf{Setup.} We evaluate on 20 KramaBench tasks sampled across domains: 10 runtime-dependent tasks and 10 control tasks without runtime bindings. Eight of the runtime-dependent tasks are \emph{value-dependent}, where a computed value or entity list selects members of a table collection. In the other 2, the binding is a data format observed only after inspecting an initial table.

We test Gemini 3.8 Flash and Claude Sonnet 5, each serving as both \name and the host agent. \name uses the catalog and attached metadata as described in Section~\ref{sec:exp_setup}, with up to 20 LLM calls per invocation. The host agent runs SQL and Python in a sandbox restricted to the task's dataset. It never exhausts the total budget of 40 LLM calls per task.

We report two metrics, averaged over three runs. \emph{Task accuracy} scores the host agent's final answer with the metrics reported in the original paper~\citep{lai2025kramabench} after normalizing case, punctuation, and units. \emph{MR F1} scores \name's final handoff against the tables required once the runtime binding is known on runtime-dependent tasks, and against the ground-truth tables on control tasks. All the results are in Table~\ref{tab:integration_e2e}.

\begin{table}[tb]
\centering
\caption{Task accuracy (\%) and MR F1 (\%, mean over 3 runs; $\pm$ is task-accuracy std.\ dev.). Bold marks the best strategy.}
\label{tab:integration_e2e}
\resizebox{\linewidth}{!}{
\begin{tabular}{lrrrr}
\toprule
 & \multicolumn{2}{c}{All tasks (20)} & \multicolumn{2}{c}{Runtime-dep. (10)} \\
\cmidrule(lr){2-3} \cmidrule(lr){4-5}
Strategy & Task Acc. & MR\ F1 & Task Acc. & MR\ F1 \\
\midrule
\multicolumn{5}{l}{\textbf{Gemini 3.8 Flash}} \\
S1: Single-pass            & 72.8 {\scriptsize$\pm$5.0} & 77.1 & 79.0 {\scriptsize$\pm$5.8} & 79.9 \\
S2: Host-agent-driven      & 75.3 {\scriptsize$\pm$2.5} & 77.6 & \textbf{85.7} {\scriptsize$\pm$0.0} & 84.6 \\
S3: Discovery-driven       & \textbf{77.2} {\scriptsize$\pm$1.2} & \textbf{80.6} & 84.3 {\scriptsize$\pm$2.3} & \textbf{84.7} \\
\midrule
\multicolumn{5}{l}{\textbf{Claude Sonnet 5}} \\
S1: Single-pass            & 73.7 {\scriptsize$\pm$3.8} & 73.2 & 84.0 {\scriptsize$\pm$5.8} & 75.9 \\
S2: Host-agent-driven      & 73.7 {\scriptsize$\pm$1.4} & 76.0 & 82.3 {\scriptsize$\pm$7.6} & 77.6 \\
S3: Discovery-driven       & \textbf{74.8} {\scriptsize$\pm$3.0} & \textbf{77.8} & \textbf{86.3} {\scriptsize$\pm$4.5} & \textbf{79.2} \\
\bottomrule
\end{tabular}}
\end{table}

\noindent\textbf{Exchanges improve table selection.}
S3 achieves the highest MR F1 on both models. On runtime-dependent tasks, it exceeds S1 by 4.8 percentage points on Gemini 3.8 Flash and 3.3 points on Claude Sonnet 5. This improvement comes from returning fewer unneeded tables to the host agent: on value-dependent tasks, S3 reduces the average number of selected tables from 6.5 to 4.4 on Gemini 3.8 Flash and from 7.7 to 4.5 on Claude Sonnet 5. For example, in the wettest-summer task, S1 returns all 23 annual water-testing tables, whereas S3 requests the wettest year and selects only that year's table.

S2 rarely completes an exchange, which requires \name to defer the dependent collection and the host agent to re-invoke \name. Of the 30 runtime-dependent runs per model, \name defers in only 6 on Gemini 3.8 Flash and 3 on Claude Sonnet 5. S3 completes an exchange in 11 and 9 runs, respectively, and matches or exceeds S2 on runtime-dependent MR F1.

\noindent\textbf{More precise table selection yields more accuracy gains.}
S3 also achieves the highest overall task accuracy on both models, exceeding S1 by 5.3 points on Gemini 3.8 Flash and 2.3 points on Claude Sonnet 5 over runtime-dependent tasks. The gains are largest on value-dependent tasks (6.6 and 5.0 points, respectively). The gain is small because the host agent often compensates for an imprecise handoff: under S1, it finds the required member itself in 87\% or 92\% of value-dependent runs on Gemini 3.8 Flash and Claude Sonnet 5, respectively. We expect the accuracy gain to grow with collections too large for the host agent to explore.

\noindent\textbf{Exchanges add little cost.}
Summed over both agents, S3 uses 9\% more tokens per task than S1 on Gemini 3.8 Flash (797k vs.\ 728k) and 3\% more on Claude Sonnet 5 (348k vs.\ 339k), an overhead from serving execution requests.

%% file: sections/7_conclusion.tex
\section{Conclusion and Future Work}

As agents take on enterprise analytics and data engineering workflows, their main bottleneck is shifting from generating code to grounding their reasoning in the right data sources across messy data lakes. 
This paper argued that one-shot similarity retrieval alone cannot select a sufficient and minimal set of datasets, and that this task requires metadata reasoning as a first-class agentic capability. 
\name turns passive, human-centric data catalogs into active context layers for AI agents. 
It matches discrimination-oriented search, attached table summaries, and on-demand specialized tools to each step of discovery, and interleaves discovery with query execution when table selection depends on runtime values. Deployable as a standalone catalog service or a lightweight harness plugin, \name offers a practical foundation for production data agents.

Our work opens three research directions at the intersection of data management and agentic AI. First, \emph{self-evolving data catalogs} could close the loop between agent execution and metadata curation. As agents verify join paths, uncover silent data-quality issues, and resolve runtime bindings, \name could write these findings back to the catalog to enrich lineage, flag deprecated tables, and fill in missing business glossaries. Second, metadata reasoning can extend from relational tables to \emph{heterogeneous, federated data estates} spanning unstructured documents, knowledge graphs, external APIs, and live streams, where selecting a minimal set of sources must balance semantic coverage, data freshness, access control, and query cost. Finally, \emph{co-optimizing discovery and execution agents} could let systems trade fidelity for latency, for example by choosing between approximate pre-aggregated views and fine-grained base partitions based on downstream SLAs.